# DFT modeling of the covalent functionalization of graphene: from ideal to realistic models

D. W. Boukhvalov

*School of Computational Sciences, Korea Institute for Advanced Study, Seoul 130-722, Korea.*

*The production of multiple types of graphene, such as free standing, epitaxial graphene on silicon carbide and metals, graphene in solution, chemically grown graphene–like molecules, various graphene nanoribbons, and graphene oxide with different levels of reduction and various chemical composition, demonstrate the need for additional investigation beyond the basic principles of graphene functionalization for avoidance of occasionally contradictions between the predictions from first-principles simulations and experimental results. Herein, we discuss the current state of modeling of the different types of graphene using density functional theory (DFT) methods. We focus on the static (substrate, shape, curvature, strain and doping) and dynamic (starting point of functionalization, migration barriers and stability of configurations) aspects that provide a more correct and selective modeling of the chemisorption of various chemical species on the graphene scaffold. Based on the recent modeling of experimentally realized functionalization of different types of graphene we can conclude that the formation of uniform one- or two-sided functionalized graphene discussed in earlier studies is an exception to the typical scenarios of graphene chemistry. The presence of different substrates, defects and lattice distortions, such as ripples and strain, results in the formation of clusters or lines from the functional groups. Several configurations of the chemical species on the graphene substrate have been found to exist with ideal models but are only stable for graphene functionalized under special conditions. The role of the technical parameters, such as the functionals and computational methods employed, and the important guidelines for the modeling of the chemical properties of various types of graphene are also discussed.*

E-mail: danil@kias.re.kr

# 1. Introduction

The discovery of graphene[1] attracted much attention from the physical community. After the first reports on the interesting and promising properties of graphene oxide[2] and the potential employment of this material for large-scale graphene production,[3] it has become the subject of intensive experimental and theoretical chemistry studies. It is not intuitive for a one-atom-thick membrane that has a surface but not bulk should have unusual chemical properties. The prediction of the transformation of semi-metallic graphene to insulating 100% hydrogenated graphene (graphane) had been predicted by theory[4,5] and proven experimentally.[6] In addition, a description for restoring the conductivity of graphene oxide after significant reduction[7] has been confirmed experimentally[8] demonstrating that density functional theory (DFT) is a powerful and useful tool for describing and predicting the chemical and physical properties of pure and functionalized graphene.

The hydrogenation of graphene is the simplest chemical reaction for this compound, and this reaction is simple and comprehensive enough for a discussion of the basic principles of graphene chemistry.[9] In our first review on this subject, we discussed the three main principles for the hydrogenation of graphene.

- Broken bonds are energetically unfavorable, resulting in magnetic states on graphene being unusually unstable.
- Graphene is very flexible, and atomic distortions strongly influence the chemisorption process.
- The most stable configurations correspond to 100% coverage for two-sided functionalization and 25% coverage for one-sided functionalization.

The early goals of graphene chemical functionalization (e.g., manipulation of energy gap, graphene oxide reduction, production of graphene and protection of graphene from oxidation) were modified by the appearance of novel graphene based systems, such as graphene – boron nitride layered structures[19] or graphene oxide – metal oxide composites. Novel methods for graphene production, such as growth on metal substrates,[20] and novel types of graphene doping, such as insertion of nitrogen (see references in [21]), boron[22] and sulfur[23] atoms in graphene membranes, significantly expands the challenges for computational chemistry. The discovery of catalytic properties of graphene oxide[24,25] and nitrogen-doped graphene designates graphene as a prospective material not only for use in future electronics but also for use in green chemistry. These novel types of graphene and its applications for various chemical reactions require development of adequate and unique models for each type of graphene and each chemical process. Additionally recent experimental works[26,27] demonstrates significant influence of the covalent functionalization to the transport properties of graphene. In the theoretical works discussed this aspects[28-34] has been discussed only single or randomly distributed or single chemical species on graphene. Theoretical works discuss enhancement of the screening,[32] backscattering,[33] transport gap opening[28] caused by the impurities. The role of the distribution of impurities for the electron-hole asymmetry[32] and transport gap opening[34] are also pointed in the theoretical works. Thus chemical functionalization of graphene can be also the source of manipulation of the transport properties of graphene based devices and its parts. Obtaining of the correct model of atomic structures of experimentally fabricated of proposed samples of functionalized graphene is the first and unavoidable step for the further description of its electronic and transport properties by the using of the results of DFT calculations within the model approaches[28-33] or direct transport calculations.[34] In this brief account, we present a discussion regarding the transformations of general models and principles for application to different realistic types of graphene with a discussion including several important tips and tricks.

## 2. Methodological aspects of the modeling of covalent functionalization of graphene

The primary options required for the modeling of graphene chemical functionalization are implemented in all broadly distributed computer codes. The choice of the proper functional is very important for the correct description of the energetics of covalent functionalization (see Table I). For the configuration of adsorbed species, the choice of the appropriate functional is not important, except for the layered systems (see below). As we previously demonstrated, GGA-PBE[35] provides the correct estimation of the chemisorption energy for the hydrogenation[6] and oxidation[9,21] of graphene.

For calculation of the correct energy gap value, a more accurate GW approach should be employed instead DFT.[36] The primary difficulty with the use of this method for functionalized graphene is the high computational cost. Currently, only systems with 20 or fewer atoms per unit cell can be calculated without extraordinary software and hardware requirements. However, this limitation is not critically important for the modeling of realistic graphenes because the exact calculation of the energy gap could be performed only for cases with high coverage levels (above 12 atomic % of carbon – 1/8 carbon atoms are functionalized) that correspond to unit cells with fewer than twenty atoms. An estimate of the order and value of the experimentally measured[37] energy gap is sufficient from a theoretical standpoint for the discussion of the experimental results. The other routes for the solution of the energy gap problem is employment of hybrid functionals[38] or the methods of the fitting of energy gap value for the large supercell.[39]

**Table I** Comparison of the main properties of the computational methods and functionals.

| Functional or method | Binding energies | Interlayer distances | Energy gap | Notes |
|---|---|---|---|---|
| GGA | correct | overestimation (without dispersion forces correction) | underestimation | Implemented in all codes |
| LDA | overestimation | correct (due to error cancellation) | underestimation | Implemented in all codes |
| Hybrid | correct | correct | correct | Special choice of the functional is required. Implemented in several codes |
| GW | correct | correct | correct | Below 20 atoms can be calculated |

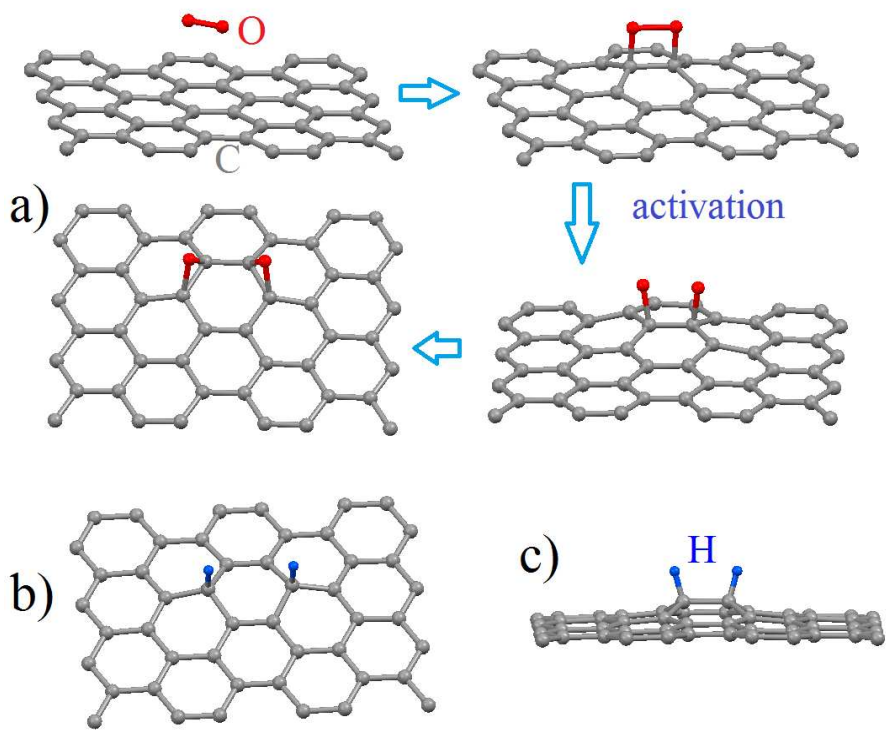

**Figure 1** Optimized atomic structure of step-by-step chemisorption of molecular oxygen (a) and formation of pairs of hydrogen atoms in para (b) and ortho (c) positions on graphene.

Chemisorption energy of the various species on graphene are calculated by a standard formula $E_{chemisorption} = (E_{graphene\ with\ chemisorbed\ species} - E_{graphene\ before\ adsorption\ this\ species} - E_{species})/N$, where the $E_{graphene\ with\ chemisorbed\ species}$ and $E_{graphene\ before\ adsorption\ this\ species}$ the total energies of the pure and functionalized graphene and $E_{species}$ - the chemical species in an empty box are employed, and N is the total number of adsorbed species. This value does not include the activation energy (energy required for the dissociation of covalent bonds within the species, see Fig. 1a) of the chemical species or the weak forces between the species in liquid or gas phase. These weak forces are estimated to be lower than 0.4 eV/specie in the liquid and 0.1 eV/specie in the gas phase and can usually be omitted from the discussion because the energy required for the dissociation of these bonds is lower than the energy required for graphene functionalization.[6] The activation energies can be comparable to the values of the chemisorption energies[21] and are strongly dependent on the local chemical properties of graphene. For example, in the vicinity of nitrogen impurities, the activation energy of the oxygen molecule can decrease by approximately three times.[21] Similar to the employment of graphene for the measurement of the fine structure constant and quantum metrology, it can also be used for the justification of the ratio between the DFT calculated formation energies and the experimental temperatures because graphene is rather clean and the chemical processes are simple. The calculated value of the activation and chemisorption of oxygen on pure graphene is approximately 1.4 eV[21] and the experimentally measured temperature is approximately 200 °C.[13] The graphane dehydrogenation temperature is approximately 400°C,[6] which corresponds to the calculated single vacancy formation energy that is more than 2 eV. Based on Boltzman formula which is standard for the description of the defect formation[40,41] and chemisorption of species on the carbon substrate[42] $C=C_0 e^{(-E_{chem}/k_B T)}$, where C is the probability of chemisorption, $k_B$ – Boltzman constant, T – temperature of the process and $C_0$ – specific constant for the material we can predict the rate of chemical processes over graphene substrate. Based on obtained above calculated energies and experimentally measured temperatures we can predict that for realizing of the processes with chemisorption energy above 3 eV with high rate temperatures above 700°C is required.

Formation energies describe only static part of the functionalization process. But dynamical aspects such as migration barriers of ions[24,43] or molecules[44] over graphene substrate and accessibility of active sites of graphene[45] occasionally can provide the valuable changes in the reactions pathways. Usually migration barriers for the ions during functionalization process are about 0.3 eV. This value is lower than usual formation energies of the processes (see detailed discussion in Ref. [24]). Energy cost of reactions could be increased for the functionalization of graphene by the molecules in gaseous phase due to formation of strong non-covalent (π-π in the case of aromatic molecules or ionic in the case of bromine)[43,46] bonds between graphene and species. The changes of the accessibility of the graphene plateaus due to decreasing of the interlayer distances in layered graphene based structures are also increase the energy costs of functionalization processes and makes these processes impossible at for the used temperatures (see detailed discussion in Ref. [45]). Hence for the correct description of the energetics of the graphene functionalization all possible sources of the increasing of the energy cost of the studied reactions should be checked.

Another important issue for the discussion of energetics of graphene functionalization is the underestimation of weak bonds when the GGA functional[47] is employed, which is important for an exact description of layered systems beyond the bilayer.[48] Additional corrections to account for van der Waals and similar forces are implemented in several codes. LDA[49] reproduces the interlayer distances in graphite rather well[28] due to error compensation. However, this approach results in an overestimation of the binding energy of covalent bonds.[5] If this correction does not implemented in used code simple method for the correction of chemisorption energies within LDA can be used.[51] At the end of the methodology chapter, I would like to suggest useful method for significantly accelerating the DFT calculations of the functionalization of graphenic systems with large number of atoms in the supercells. First, obtain a reasonable level of self–consistency for a single k-point and the minimal energy value for the mesh cut–off required for the convergence of the iterative processes. Then, further calculations from the optimized electronic and atomic structure can be performed with the appropriate technical parameters that provide accurate results. A combination of this method along with taking into account the basic principles of graphene functionalization can drastically improve the speed of the calculations and allows for a larger numbers of atoms per supercell to be studied.

3. Models of realistic graphenes

3.1. Step-by-step functionalization and role of starting points and lattice distortions

3.1.1. Role of starting point

In the early studies, graphene functionalization[4,5,7,51] has been discussed in terms of the chemisorption of a single adatom and pairs of impurities as well as the ground state and most energetically favorable intermediate configurations. Further modeling demonstrated that the route to energetically favorable final configuration often proceeds through energetically unfavorable intermediate steps.[9,44] The primary cause of this variation in the energetic favorability is the distortions in the flat graphene. The energetically favorable configurations usually correspond to diamond–like structures. The close proximity of undistorted non-functionalized and distorted functionalized areas can be very energetically unfavorable, and the addition of an additional atom or pair of chemical species can enhance the attraction between this two areas. Another source of these energetically unfavorable intermediate steps involves the noncoincidence of the two-fold symmetry of graphene lattice distortions created by the Diels-Adler[52] addition of a pair of atoms or species and the symmetry of nano-graphenes or graphene edges.[45] A simple example of these effects is the limitations of the reduction of large polycyclic aromatic hydrocarbons discussed in Ref. [45] (Figure 2). An example of the crucial role of step-by-step functionalization is the hydrogenation of the graphene bilayer. The most stable configuration is the passivation the sublattice of one layer from the top side and the other sublattice of the other layer from the bottom side by the formation of carbon-carbon covalent bonds between the graphene layers resulting in magnetism and semiconductivity.[53] However, the formation of this structure is rather difficult because the first steps leading to the adsorption of the hydrogen pair in the para position (Fig. 1b) is much more energetically favorable than in the meta position, which is the first building block in the desired structure. Thus, the favorability of the initial state will provide a lower level of hydrogenation (only 25% instead of 50% of the carbon atoms will be hydrogenated) with the formation of a nonmagnetic compound with a different band gap value.[48]

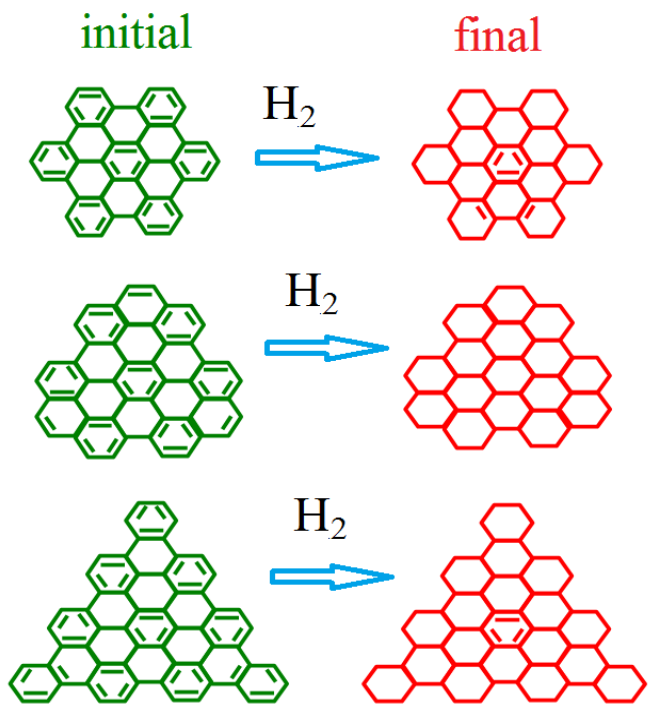

**Figure 2** Structures of realistic policyclic aromatic hydrocarbon before and after exposing by molecular hydrogen.

In addition, the starting point of the functionalization is important for the description of functionalization aspect. The presence of defects, such as vacancies, substitution of a carbon atom by another, and Stone-Wales or other similar defects, can dramatically reduce the value of the chemisorption energy resulting in chemisorption at the defect sites, which occurs for the graphene edges.51 If the adsorption occurs initially on the edge, it can dramatically change the scenarios of the graphene functionalization. For example, for graphene reduction by molecular hydrogen in the presence of a defect, chemisorption of the first hydrogen pair on this defect is energetically favorable and each successive functionalization step requires more energy. Therefore, a hydrogenated cluster in the vicinity of the defect will be formed. For perfect graphene, one-sided hydrogenation results in the formation of uniform coverage by decreasing of the chemisorption energy after each successive step of the hydrogen pair adsorption (Fig.

3). The difference between the functionalization scenarios of perfect and imperfect graphene is similar to two-sided hydrogenation on the same systems.[45]

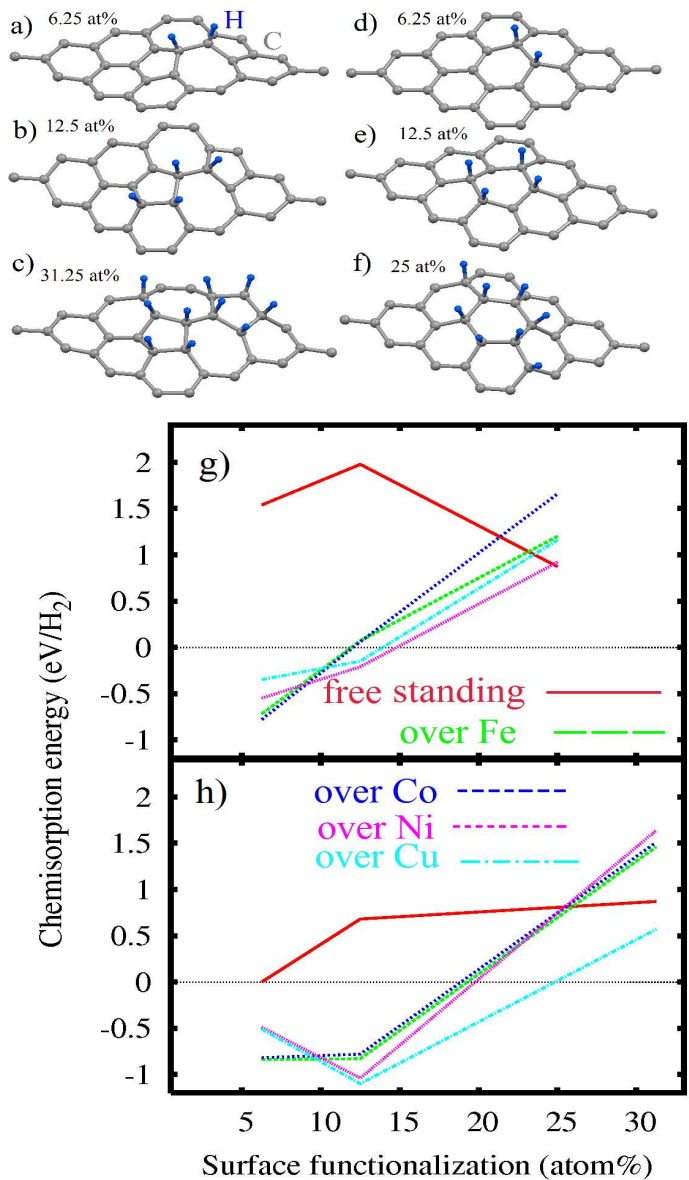

**Figure 3** Optimized atomic structure of step-by-step hydrogenation of perfect (a-c) and imperfect (e-f) graphene and chemisorption energies of these proces as function of coverage and substrate (g, h).

### 3.1.2. Role of chemical composition of the species on graphene

The modeling of the functionalization processes must take into account the chemical composition of the species. For example, the reduction of graphene by hydrogen plasma results in the adsorption of the first pair of hydrogen atoms in the most energetically favorable para position (Fig. 1c).[48] However, the use of molecular hydrogen for these reactions results in the formation of the first pair of adatoms in the ortho position.[45] Chemisorption of the fluorine goes by the similar way with formation similar patterns and lager carbon-adatom distances, lattice distortions and binding energies.[55] For graphene oxidation, the source of oxygen (mineral acids, oxygen plasma, ozone or molecular oxygen) is important for discussing the formation of the first pair of oxygen atoms and correctly determining the activation energy of molecular oxygen.[21] The hydrogen bonds between hydroxyl groups could also play a role in the graphene functionalozation.[56-58]

### 3.1.3. Role of local distortion of graphene flat

In early models of graphene functionalization, only the perfectly flat membranes were employed. This model was adequate for the current view of graphene as a planar two-dimensional system. Further discovery of the intrinsic and substrate induced ripples in graphene (see Ref. [59] and references therein) resulted in a series of studies elucidating the roles of the ripples in determining the chemical properties of graphene.[59,60] The main source of this chemical activity is from the rotation of the π-orbitals, which has also been reported for carbon nanotubes[55] and the formation of mid-gap states in the electronic structure of enormously corrugated graphene.[53] The relatively flat ripples in epitaxial graphene and the graphene bilayeras well as the larger intrinsic ripples in free standing graphene provide minor enhancement of the chemical activity, and only the large corrugations caused by quartz or metal substrates and bending graphene[62] lead to a drastic increase in the binding energy and the formation of stable clusters of adatoms,[59] or lines.[60] Similar to the out-of-plane distortions (ripples), in-plane compression or expansion of the graphene sheet (strain) leads to enhancement of graphene's chemical activity due to the increase in the total energy of the distorted graphene. Similar to the deviation of uniform functionalization of graphene, this distortion could be the source for the variation in the functionalization scenario. For example, the lines of oxygen atoms along the zigzag directions previously discussed for nano-graphenes (see below) can be induced in flat infinite graphene only by strain in the zigzag direction.[54] All discussed effects should be taking into account for the further development of the approaches for descriptions of large scale functionalization such as cluster expansion method.[55]

## 3.2. Size of supercell and stability of magnetic configurations

### 3.2.1. Role of supercell size for the correct description of atomic structure

For modeling of graphene chemical properties, two different model of the graphene sheet can be employed. The first model is the supercell of infinite graphene within periodic boundary conditions, and the second model involves a large nano-graphene molecule containing at least 24 carbon atoms in an empty box. The choice of model is often determined by the approach and computer code employed. For computational codes based on quantum chemical methods, only the second approach is viable. There are three primary effects caused by the size of the supercell or the nano-graphene molecule. The first effect results in variation of the chemisorption energy from distance between the species on graphene scaffold.[5] The second effect involves the adsorption of the adatoms pairs in different configurations. The third effect is significant changes in the migration energies.[57] All of these effects are caused by the distortion of a large graphene area by the chemisorption of chemical species. For example, the chemisorption of a single hydrogen adatom or pair of hydrogen atoms results in strong out-of-plane corrugation of the graphene sheet with a magnitude ranging from 0.22 to 0.1 Å within a radius of approximately 5 Å (two lattice parameters or four coordination shells of carbon atoms connected to hydrogen) and non-negligible corrugations within a radius of 1 nm (four lattice parameters and eight coordination shells).[7] A small overlap of these distortion areas does not significantly change the chemisorption energy,[5] and the estimation of this value requires a minimum supercell size consisting of 32 carbon atoms. In contrast to chemisorption of the first chemical specie, the adsorption of the second impurity should be strongly dependent on the size of the supercell or nano–graphene. An example of the variation of chemisorption patterns with increasing nano-graphene size is shown on Fig. 3. For smaller nano-graphene, the formation of lines of epoxy-groups along the zigzag direction, which has been proposed as a source of graphene unzipping,[64] are much more energetically favorable than the formation of other types of oxygen atom pairs. The primary cause of this change in the pair formation scenarios is the decay in the flexibility of larger nano-graphene samples (Figs. 4b,c) that favor the chemisorption of pairs resulting in the lowest graphene sheet distortion. The rigidity of larger nano-graphenes results in strain that is less energetically favorable along the zigzag direction, which is necessary for the formation of oxygen lines.[64] Inducing of corrugation by the folding of graphene into single wall carbon nanotubes[65,66] or by the formation of oxygen clusters[67] also leads the formation of the oxygen lines along zigzag axis. Therefore, for the modeling of the early stages of chemical functionalization, larger (approximately 72 or more carbon atoms) supercells or nano-graphenes are preferable. Enlargement of the nano-graphenes results in a change in the functionalization scenario when the formation of the described oxygen lines becomes less energetically favorable than other configurations.

The overlap out-of-plane lattice distortion areas are also important for the migration of chemical species over the graphene substrate.[57] For different chemical species, the variation in these barriers could be different. In addition, different calculation methods and the size of supercell used in the calculation results in different migration barriers (see references in [57]). In general, the migration pathways are calculated using the nudged elastic band (NEB) method.[68] However, for graphene, this method can be simplified.[57] For example, the migration barrier for hydrogen atoms, which cause smaller out-of-plane graphene sheet distortions, increase as the distance between the hydrogen impurities increases. Therefore, a supercell with more than 72 atoms is required for the correct description of the migration of chemical species over flat graphene.

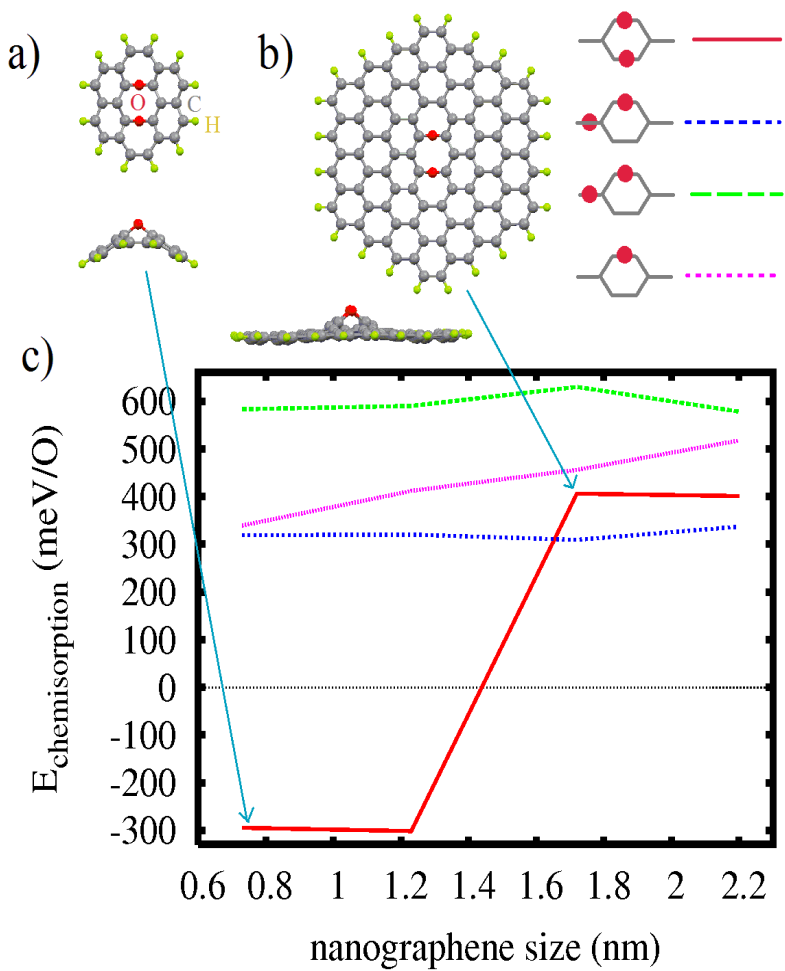

**Figure 4** Formation of oxygen patterns on coronene-like nanographenes as function of its size.

### 3.2.2. Stability of magnetic configurations

This type of magnetism created by unpaired electrons from dangling bonds appears after the chemisorption of an odd number of mono-valence chemical species[5,69,70] or after the formation of zigzag graphene edges[69,71,72] and is localized on the graphene sublattice. The chemisorption of additional species on this sublattice or subsequent migration of other adatoms results in the disappearance of the magnetism due to passivation of the dangling bonds. The large migration barriers of the hydroxyl groups and its large size render it a candidate for the creation of *sp*-magnetism in graphene.[70,72] The observation of magnetism in graphenic systems with unavoidable[16,18] or the possible[15] presence of these species is indirect confirmation of this experimental results. The other important issue related to the sp-magnetism in graphene systems is the contention between the structural stability and the existence of exchange interactions sufficient for the magnetic order. To illustrate this aspect of sp-magnetism, two examples will be discussed. The first example is graphone,[73] which is ferromagnetic graphene with a single hydrogenated sublattice. The results of additional calculations have shown that the lattice should be converted by a recombination of hydrogen atoms to the non-magnetic configuration.[74] Another example is the enhancement of ferromagnetic interactions between large-spin magnetic clusters as the distance decreases between them resulting in a decrease in the migration barriers for the reconstruction of magnetic clusters to non-magnetic.[42,71] This disagreement between the stability of structure and the existence of intrinsic ferromagnetism is a challenge for computational chemistry. Currently, several potential solutions have been proposed from theoretical investigations including:

- to employ chemical species with larger migration barriers,[42,70,74]
- to topologically stabilize the impurities with lattice defects, such as Stone-Wales defects,[51] ripples,[71] graphene lattice compression[57] or competition between *sp2* and *sp3* hybridized areas[77]
- to convert weak antiferromagnetic interactions between localized spins to strong ferromagnetic interactions via injection of additional electrons into system.[76-78]

### 3.3. Substrate effects

### 3.3.1 Substrate induced ripples

In contrast to quartz or other weakly bonded systems employing van der Waals forces, a number of other substrates with much stronger graphene-substrate bonds exist. The simplest example of graphene that is relatively strongly bound to the substrate is multilayer graphene and graphite. Spectroscopic measurements demonstrate that a multilayer graphene with more than 6 layers is indistinguishable from graphite. Therefore, six-layer graphene can be employed as a good model of graphite.[79] The functionalization of the graphene multilayer surface can be discussed in terms of the adsorption of species over graphene on multilayer graphene. There are two specific points for the covalent functionalization of multilayer graphene. The first issue is the absence of substrate buckling and other inhomogeneities, which provides an opportunity for uniform functionalization. In addition, the effect of ripples does not need to be addressed. The second issue related to this type of substrate is the decrease in the graphene flexibility due to the interlayer interactions resulting in variation of the chemisorption energies.[5,71]

### 3.3.2. Silicon carbide substrate

The other type of strongly bonded substrate is silicon carbide. Recent experiments have reported different hydrogenation scenarios for epitaxial graphene[10,11] compared to theoretical prediction.[5,48] The unexpected equal propagation of hydrogen pairs in the ortho and para positions was in contradiction to the theoretical prediction of energetically favorable hydrogen pairs in the para position[5] and the formation of stable hydrogen[11] clusters instead of the predicted uniform coverage.[25] The presence of the unavoidable buffer layer between carbon layer(s) and the insulating substrate results in variation in the interaction between the graphene layer and the scaffold. The exact description of the electronic structure of epitaxial graphene requires taking into account several atomic layers of the substrate, the face of substrate and the buffer layer as well as the mismatch between the graphene and the silicon carbide lattice parameters.

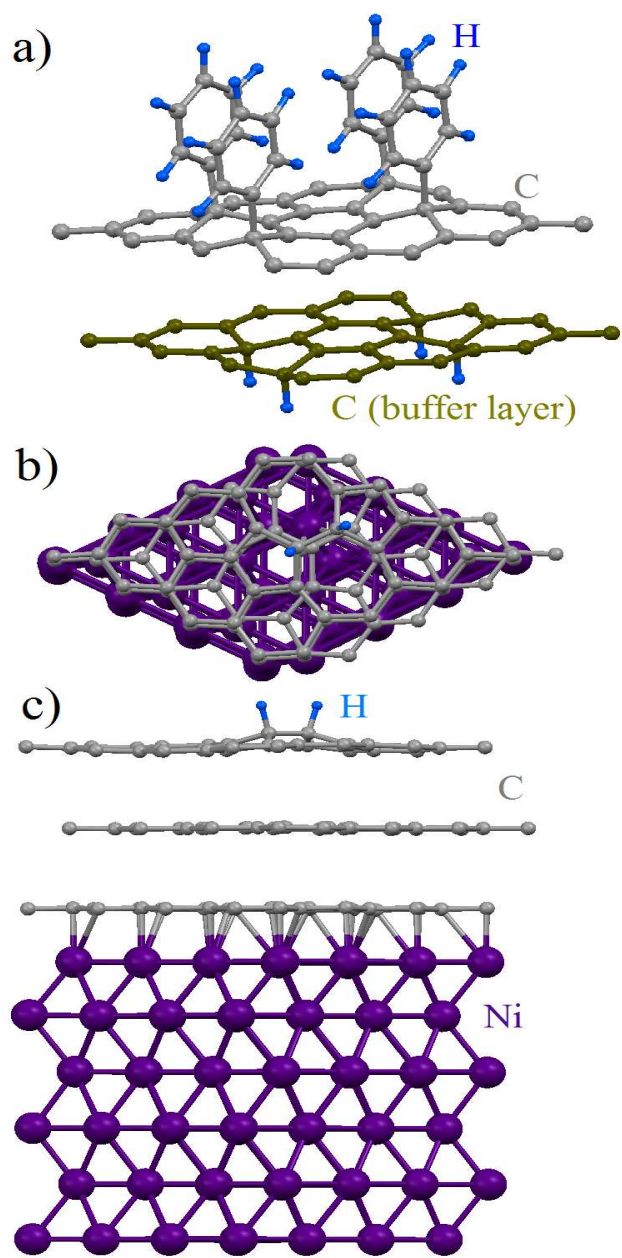

**Figure 5** Optimized atomic structure of the model of epitaxial graphene with chemisorbed phenyl groups (a), and top (b) and side (c) views of imperfect graphene on Ni(111) with chemisorbed pair of hydrogen atoms.

Fortunately, a simplified and computationally cheaper approach can be employed to model the chemisorption of various species on this type of graphene.[80] Within this model, the substrate and buffer layers can be substituted by a partially hydrogenated second graphene layer with a hydrogen coverage below 25% (see Fig. 5a), which significantly decreases the number of atoms in the supercell and permits the calculation of the chemisorption of rather large species on the surface of epitaxial graphene. This approach has been able to explained all of the experimentally obtained results, such as the formation of a nearly equal number of hydrogen pairs in the ortho and para positions at low hydrogen coverage,[10,11] the formation of stable clusters of hydrogen,[11] the appearance of the energy gap and the maximum coverage level after arylation (Fig. 4a),[25] and the penetration of the lithium ion throughout the epitaxial graphene membrane at room temperature.[81]

### 3.3.3. Metals substrate

Another type of strongly bonded substrate involves various metals (Fig. 4b,c). Since the pioneering work that reported the exfoliation of graphene from a nickel substrate,[20,82] carbon vapor deposition has been discussed as the best route for industrial manufacturing of graphene. The prediction of intriguing physical properties of graphene on metal substrates[83] makes these compounds as interesting as graphene itself. Within the last few years, various transitional metals have been employed for the growth of graphene. The metals can be divided into two categories as follows: *3d* metals and *4d* metals. The special characteristics of *3d* metal substrates include the disappearance of the mismatch between the metal and graphene lattice as well as the strong electrostatic interactions between the graphene and the metal substrates.[84] The electrostatic interactions between graphene and the *3d* metal substrate are weaker than in the carbides of these metals but stronger than between graphene and the *4d* and *5d* substrates where the carbon-metal distances are approximately 2.5 Å smaller than the typical van der Waals distance (approximately 3.5 Å) and larger than the distances in transitional metal carbides

(approximately 2.1 Å). The properties of the graphene *3d*-metal interface result in the formation of nearly flat graphene in contrast to the puckering observed for graphene over *4d* metals.[85,86] The presence or absence of this puckering provides graphene functionalization scenarios similar to the functionalization of flat and rippled graphenes.

Another important chemical property of graphene results from *n*-type doping from metal substrate.[83] This injection of electrons provides enhancement of the chemical activity in graphene[87] and the decrease in the energy required for the formation of defects.[88,89] An increase in the strength of the graphene–metal electrostatic interactions correlates with the enhancement of graphene's chemical activity. To illustrate this effect, the initial hydrogenation steps of graphene over various *3d* metals have been performed. The presence of the 3d metal substrate converts the high positive value for chemisorption energy to a small negative value (Fig. 3g,h). In contrast to uniform functionalization of perfect free standing graphene corresponding to a decrease in the formation energy with the coverage growth, the chemisorption energy increases in the presence of the metal substrate, which corresponds to the limitation of the functionalized area (formation of clusters). The chemical activity of graphene over metal substrates is also affected by buckling, which has been experimentally observed in graphene grown on a cooper substrate[62] or caused by the Stone–Wales (Fig. 5b,c) and similar defects that are at current time unavoidable in graphene on various transitional metals,[88] and typically caused by the imperfect stacking between different graphene domains, the initial defects in the graphene growth areas or resulting from the interaction with the transition metal substrate[88] and distortions of the metal substrate.[89] The functionalization of the area near Stone-Wales defects in graphene on a metallic substrate is more favorable than functionalization of free-standing graphene (Fig. 3g,h) that correspond with the oxidation of this material at ambient conditions.[89]

# 4. Manipulation of band gap values insights from first-principles calculations

For the obtaining of reasonable value of the band gap (within 1 eV) and survival of high charge carriers mobility uniform coverage is necessary.[78] Form DFT calculations we can discuss a proper routs toward this goal. Functionalization of free standing insignificantly corrugated graphene with employment of molecular hydrogen or fluorine should provide limited coverage of graphene sheets according to the results discussed for nano-graphenes.[45] Step-by-step modeling of this process for the rather large graphene sheets and different concentration of gaseous species was not performed at this time. Results of this modeling can be calculated realistic patterns of hydrogen or fluorine of graphene and suggestions for the experimentalists regarding desirable concentration of used species. Functionalization of free standing graphene with employment of larger molecules is extremely sensitive to the corrugation that lead formation of the spot like patterns in out of flat graphene areas.[90] Searching of the chemical species that should not form spot like patterns and provide uniform functionalization with coverage below 100 at% on any corrugations of free standing graphene flat is also subject of the further theoretical works.

Another rout of graphene band-gap engineering is functionalization by atomic species. As suggest theory[4,5] and experiments[6] 100 at% coverage should be achieved. Other theoretical works (see Ref. 39 and references therein) suggests probable uniform patterns suitable for the formation of lower band gap. There is two unresolved problems in this area (i) is step-by-step description of the formation of these or other patterns from atom-by-atom addition to graphene and (ii) is the large value of the minimal energy gap. In recent theoretical works[5,39] band gap opened from the hydrogenation level above 60 at% and value of calculated energy gap above 1 eV. If take into account underestimation of the energy gap values in DFT method obtained in discussed works value is bigger than required. Probable way of solution of this problem is calculations for larger supercells or modeling of treatment of graphene with the large size radicals.

# 5. Conclusions and prospectives

The experimental results and theoretical models of realistic and proposed types of graphene have been discussed in this account provides evidence that the formation of uniform one- or two-sided functionalized graphene discussed in earlier studies is an exception to the typical scenarios of graphene chemistry. Possibility of the formation on graphene island-like structures or lines is very prospective for the fabrication of quantum-dots and nanoribbons in graphene by the covalent functionalization. The presence of different substrates, defects and lattice distortions, such as ripples and strain, results in the formation of clusters or lines from the functional groups. For the modeling of graphene on $SiO_2$ or similar compounds taking into account of substrate induced ripples is necessary, for the graphene on SiC substrate inhomogeneities of buffer layer can switch scenarios of top layer covalent functionalization, in the case of metal substrates doping and probable defects in graphene can significantly increase chemical activity of the samples and provides formation of the clusters, for the case of graphene nano-flakes and nano-ribbons the shape of the edges and size of samples is crucial for the functionalization process. The interaction between chemisorbed species and corrugated areas of graphene flat leads changes of functionalization scenarios with the changes of concentration of functional groups on graphene scaffold. Several configurations of the chemical species on the graphene substrate have been found to exist with ideal models but are only stable for graphene functionalized under special conditions. Current DFT based methods allow for computationally feasible descriptions of the chemical properties for each known type of graphene with all of the known impurities and distortions, and building of the proper models of atomic structures of realistic functionalized graphenes for the further modeling of its electronic, optic and transport properties. Realistic models of atomic structure of functionalized graphenes is also important for the further, beyond electronic, fabrication and application of modified graphene as storage material, filters, catalyst and also in area of bio-chemistry.